\documentclass[sigconf]{acmart}
\usepackage{graphicx}
\usepackage{amsmath}
\usepackage{booktabs}
\usepackage{algorithm}
\usepackage{algorithmic}
\usepackage{amsfonts}
\usepackage{multirow}
\usepackage{makecell}
\usepackage{subfigure}
\usepackage{color}
\usepackage{bm}
\usepackage{epstopdf}
\usepackage{url}
\usepackage[cal=cm]{mathalfa}
\usepackage{balance}
\usepackage{threeparttable}
\usepackage{lipsum}
\usepackage{enumitem}
\usepackage{pifont}
\usepackage{tabularx}
\usepackage{makecell}
\usepackage{float}
\usepackage[table]{xcolor}
\usepackage{array}

\setlist[itemize]{leftmargin=*}

\AtBeginDocument{%
  \providecommand\BibTeX{{%
    \normalfont B\kern-0.5em{\scshape i\kern-0.25em b}\kern-0.8em\TeX}}}

\author{Jiangxia Cao}
\authornote{Equal contribution.}
\affiliation{
  \institution{Kuaishou Technology}
  \country{caojiangxia@kuaishou.com}
}

\author{Ruochen Yang}
\authornotemark[1]
\affiliation{
  \institution{Kuaishou Technology}
  \country{yangruochen@kuaishou.com}
}

\author{Xiang Chen}
\affiliation{
  \institution{Kuaishou Technology}
  \country{chenxiang08@kuaishou.com}
}

\author{Changxin Lao \\ Yueyang Liu \\ Yusheng Huang}
\affiliation{
  \institution{Kuaishou Technology}
  \country{\{laochangxin,liuyueyang05,\\huangyusheng\}@kuaishou.com}
}

\author{Yuanhao Tian \\ Xiangyu Wu}
\affiliation{
  \institution{Kuaishou Technology}
  \country{\{tianyuanhao,\\wuxiangyu06\}@kuaishou.com}
}

\author{Shuang Yang \\ Zhaojie Liu \\ Guorui Zhou}
\authornote{Corresponding author.}
\affiliation{
  \institution{Kuaishou Technology}
  \country{\{yangshuang08,zhaotianxing,\\zhouguorui\}@kuaishou.com}
}

\copyrightyear{2026}
\acmYear{2026}
\setcopyright{cc}
\setcctype{by}
\acmConference[WSDM '26]{Proceedings of the Nineteenth ACM International Conference on Web Search and Data Mining}{February 22--26, 2026}{Boise, ID, USA}
\acmBooktitle{Proceedings of the Nineteenth ACM International Conference on Web Search and Data Mining (WSDM '26), February 22--26, 2026, Boise, ID, USA}
\acmPrice{}
\acmDOI{10.1145/3773966.3779372}
\acmISBN{979-8-4007-2292-9/2026/02}

\title{Foresight Prediction Enhanced Live-Streaming Recommendation}

\begin{document}

\begin{abstract}

Live-streaming, as an emerging media enabling real-time interaction between authors and users, has attracted significant attention. 
Unlike the stable playback time of traditional TV live or the fixed content of short video, 
live-streaming, due to the dynamics of content and time, poses higher requirements for the recommendation algorithm of the platform - understanding the ever-changing content in real time and push it to users at the appropriate moment. 
Through analysis, we find that users have a better experience and express more positive behaviors during highlight moments of the live-streaming.
Furthermore, since the model lacks access to future content during recommendation, yet user engagement depends on how well subsequent content aligns with their interests, an intuitive solution is to predict future live-streaming content.
Therefore, we perform semantic quantization on live-streaming segments to obtain Semantic ids (Sid), encode the historical Sid sequence to capture the author's characteristics, and model Sid evolution trend to enable foresight prediction of future content.
This foresight enhances the ranking model through refined features.
Extensive offline and online experiments demonstrate the effectiveness of our method.

\end{abstract}



\maketitle

\vspace{-0.2cm}
\section{Introduction}

Nowadays, live-streaming media has deeply influenced our daily lives, reshaping entertainment consumption, and redefining social interactions.
Different from traditional TV live, the video-sharing platforms like Kuaishou and TikTok have evolved the live-streaming media into a dynamic, real-time media that enables real-time engagement between live-streamers (also called authors) and users.
This transformation is particularly evident in the entertainment, where live-streaming has become a primary channel for games, music and talent shows, contributing significant commercial value.

Traditional TV live have scheduled programs at fixed times, allowing users to actively search for interested channels.
However, live-streaming platforms offer authors more flexibility for broadcast, thus there might be more than tens of thousands of live-streamings at same time, which increasing the difficulty for users to discover enjoyed live-streamings by themselves.
Consequently, to better serve both users and authors, a powerful live-streaming recommendation algorithm is required to deliver personalized content.

Over the years, models like FM~\cite{fm} and DIN~\cite{din} have formed the foundation of industrial recommendation systems. These methods perform well in content-static scenarios (\textit{e.g.}, short videos, images, music, and ads) by recommending related items based on user interaction history. 
However, as a media with ever-changing content, live-streaming is more complex to distribute to users, since they will have different experiences at different time points for the same live-streaming. 
Therefore, on the one hand, the live-streaming recommendation system needs to summarize the overall characteristics of a living room to determine whether it aligns with the users' interests; 
on the other hand, it needs to recommend live-streaming at the appropriate moment in order to provide the best experience, that is, push the live-streaming at \textit{highlight moments} to users.

\begin{figure*}[t!]
\begin{center}
\vspace{-0.2cm}
\includegraphics[width=16.5cm]{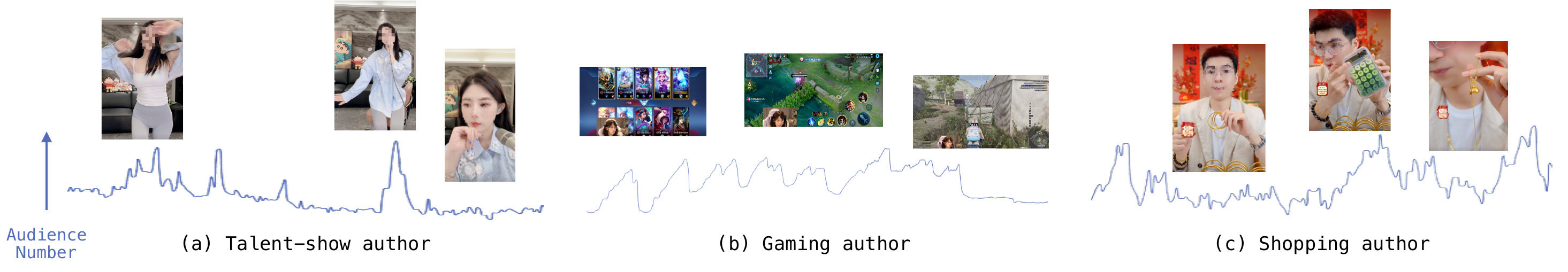}
\vspace{-0.4cm}
\caption{Cases of highlight moments in live-streamings.}
\label{fig:introfig}
\end{center}
\vspace{-0.4cm}
\end{figure*}

\textit{Highlight moments} are the key times in a live-streaming that directly showcase the author's characteristics. 
For a better understanding, here we show three cases of live-streaming interaction empirical analyses in Figure~\ref{fig:introfig}:
(i) talent-show streams see audience spikes during performances; 
(ii) gaming streams fluctuate with gameplay cycles; 
(iii) shopping streams peak in engagement when product links go live.
Based on these, we can draw a conclusion: when a live-streaming is at its \textit{highlight moments}, users will have a better experience and express more positive behaviors.
This raises a key question: how to identify and predict \textit{highlight moments}?

Fortunately, due to the short-term content continuity, a live-streaming can be divided into several semantic segments to display the content of the living room within the corresponding intervals, such as the dancing, singing, gaming, \textit{etc.}
Therefore, 
the historical process of a live-streaming can be sliced into a segment sequence,
which not only contains existing content for understanding
but also provides sufficient context for predicting the future content.

In this paper, we propose a method to enhance the performance of live-streaming recommendation through integration of historical content and foresight prediction.
Specially, by quantizing the multi-modal embeddings of live-streaming segments, we discretize the continuous content understanding into semantic ids, thereby enabling structured representation of the content. 
Previous highlight moments are contained within the historical live-streaming sequence, and indirectly modeling the dependencies among segments helps uncover these critical informational moments. 
Besides, whether the subsequent segments are highlight moments serves as a key indicator for predicting user engagement and interaction with the live-streaming. By perceiving and foreseeing such future evolution trend, it becomes possible to ensure a high-quality experience for users.
Incorporating the two types of information into the ranking model allows the model to understand the existing live-streaming content and predict users' future preferences for it. 

We conduct extensive offline experiments and analysis to demonstrate the effectiveness of segment sequence modeling and future content prediction in improving the performance of live-streaming recommendation system. The A/B testing on online platforms also shows the commercial benefits indeed brought by our method.

\section{Related Work}

Early live-streaming recommendation methods~\cite{contentctr, mmbee} are limited in key segments sampling which faces the delay in information acquisition. 
Therefore,~\cite{moment, tsstfn} continuously report fine-grained behavior slices to encourage real-time CTR prediction. 
On this basis, FARM~\cite{farm} injects cross-domain information~\cite{dmcdr} of short videos, LARM~\cite{larm} introduces LLM embeddings to enhance semantic representation.
However, these methods rely on the archived content, while the subsequent content is of greater concern to users for retention.
LiveForesighter~\cite{liveforesighter} attempts to indirectly reflect future through the generation of product sequence, but limited in intermittent content understanding.
Therefore, we aim at integrating the perception of overall future content into recommendation.

\section{Methodology}

\begin{table*}[ht!]
  \caption{Offline results in term AUC and GAUC in live-streaming ranking model.}
  \vspace{-0.3cm}
  \setlength{\tabcolsep}{7pt}
  \label{tab:offline}
  \begin{tabular}{ccccccccc}
    \toprule
    \multirow{2.5}{*}{\textbf{Model Variants}} & \multicolumn{2}{c}{\textbf{CTR}}  & \multicolumn{2}{c}{\textbf{WTR}} & \multicolumn{2}{c}{\textbf{LVTR}} & \multicolumn{2}{c}{\textbf{GTR}}
    \\ \cmidrule(r){2-3} \cmidrule(r){4-5} \cmidrule(r){6-7} \cmidrule(r){8-9} & AUC & GAUC & AUC & GAUC & AUC & GAUC & AUC & GAUC \\
    \midrule
    \multicolumn{1}{l}{\textbf{Base Ranking Model}} & 82.76\% & 64.30\% & 92.81\% & 74.24\% & 88.18\% & 73.50\% & 97.33\% & 74.24\% \\ 
    \midrule
    \multicolumn{1}{l}{\quad + \textbf{Historical Content }$\hat{I}$} & +0.00\% & +0.03\% & +0.01\% & +0.48\% & -0.01\% & +0.02\% & +0.01\% & +0.21\% \\ 
    \rowcolor{green!10}
    \multicolumn{1}{l}{\quad \quad + \cellcolor{white}\textbf{Foresight Prediction }$\hat{D}$} & \textbf{+0.05\%} & \textbf{+0.24\%} & \textbf{+0.06\%} & \textbf{+0.65\%} & \textbf{+0.03\%} & \textbf{+0.32\%} & \textbf{+0.03\%} & \textbf{+0.36\%} \\ 
    \bottomrule
  \end{tabular}
  \vspace{-0.3cm}
\end{table*}

\subsection{Segment Semantic Prediction} \label{subsec:predict}

\subsubsection{Semantic Quantization} We utilize LLM with prompt learning to achieve a multi-model semantic understanding (\textit{e.g.} author information, video, speech, comment) of live-streaming content. Due to the characteristics of live-streaming: constant change and uncertain subsequence, it is impossible to capture the content of the entire living, which also fails to meet the requirements of real-time performance. Therefore, we divide the content into 30-second segments to obtain segment semantic embeddings similar to \cite{larm}. These segment semantic embeddings strike a good balance between real-time content understanding and resource consumption. Therefore, the content of a living room can be presented through a segment semantic sequence $L_e = \{e_0, e_1, ..., e_l\} \in \mathbb{R}^{l \times d} $.

However, these high-dimensional LLM embeddings are severely dense and lack intuitive commonality, leading to expensive storage cost and difficult usage of downstream tasks. Considering the effective compression of embeddings by Vector Quantization (VQ)~\cite{vqvae} technique, which empowers the aggregation of similar semantic embeddings, we follow this practice to quantize the obtained LLM embeddings. Specifically, we implement K-Means to cluster segment semantic embeddings $\mathbf{E}$:
\begin{equation}
    \mathcal{C} = \text{K-Means}(\mathbf{E}, N), 
\end{equation}
where $\mathcal{C} \in \mathbb{R}^{d \times N}$ is the cluster centroids table, also seen as the generated codebook. 
$N$ is the number of clusters, that is, the size of codebook, and we choose $N = 20000$ to cover a wider range of categories while providing suitable quantization space.

\begin{figure}[t!]
\begin{center}
\vspace{-0.2cm}
\includegraphics[width=6.3cm]{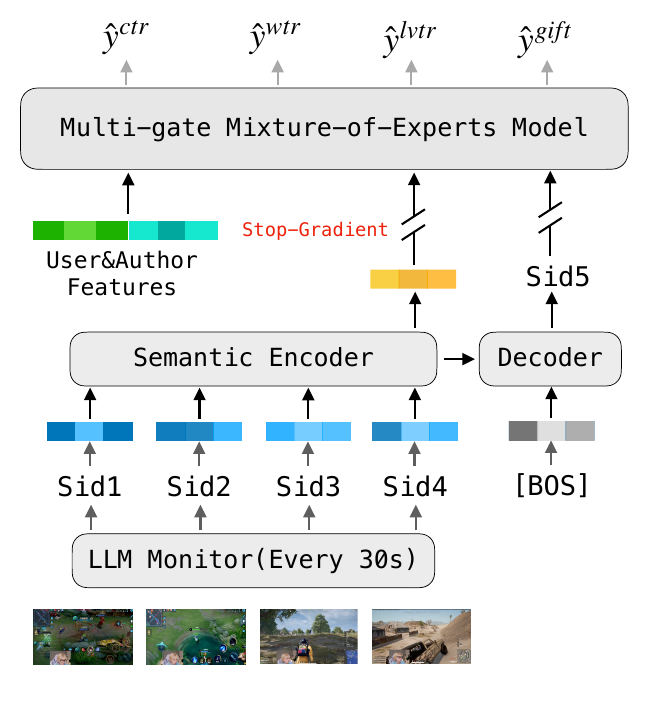}
\vspace{-0.5cm}
\caption{Overview of our model architecture.}
\label{fig:model}
\end{center}
\vspace{-0.6cm}
\end{figure}

Each segment semantic embedding $e$ can be quantized into a code $c = \text{NearestCode}(e, \mathcal{C})$, which represents a category of segments with similar content, and we regard it as \textbf{Semantic id (Sid)}. For each living room, its Sid sequence can be written as $L_c = \{c_0, c_1, .., c_l\} \in \mathbb{R}^{l}$ through quantizing $L_e$, thus achieving the transition from dense vectors to discrete codes.

However, due to the continuity of the author's content, the Sid of some intervals in $L_c$ may retain unchanged for a long time, resulting in a low information density of the sequence. To address this, we compress the sequence into two parallel sequences: 
a distinct sequence $L'_c = \{c'_0, c'_1, .., c'_{l'}\}$ that records each unique Sid,
and a corresponding frequency sequence $W_c = \{f_0, f_1, ..., f_{l'}\}$ that captures the number of consecutive occurrences of each Sid, 
where $l'$ is the length of subsequence with non-consecutive elements. This compression effectively reduces the redundancy of sequences.

\subsubsection{Semantic Prediction} Self-regressive models have demonstrates the effectiveness in generating discrete sequential tokens~\cite{tiger, onerec}, thus we conduct similar transformer architecture for the next Sid prediction. In practice, we select the last $l = 100$ segments as historical context $I = L'_c[-l:] = \{c_0, c_1, .., c_l\}$ and $W = W_c[-l:] = \{f_0, f_1, .., f_l\}$, then obtain original embedding through lookup table $I^0 = \text{EmbLookup}(\hat{\mathcal{C}} , I) + \text{EmbLookup}(\hat{\mathcal{F}}, W)$. Subsequently, the representation is processed through $l_{enc}$ layers semantic encoder:
\begin{equation}
    I^{i+1} = I^i + \text{SelfAttention}(I^i),
\end{equation}
\begin{equation}
    I^{i+1} = I^{i+1} + \text{FFN}(I^{i+1}).
\end{equation}

The semantic decoder designed to predict the next segment Sid is based on $l_{dec}$ layers cross attention mechanism. The output of semantic encoder $\hat{I} \in \mathbb{R}^{l \times d}$ are regarded as key and value, and query is the intermediate variable passed between multiple layers $D^i$, while a learnable beginning-of-sequence token serves as the query of first layer $D^0 = \{[BOS]\} \in \mathbb{R}^{1 \times d}$:
\begin{equation}
    D^{i+1} = D^i + \text{CrossAttention}(D^i, \hat{I}, \hat{I}),
\end{equation}
\begin{equation}
    D^{i+1} = D^{i+1} + \text{FFN}(D^{i+1}).
\end{equation}

The decoder result $\hat{D} \in \mathbb{R}^{1\times d}$ is the prediction embedding of the next segment Sid. We incorporate this prediction into the codebook lookup table to calculate the dot product and obtain the similarity for multi-classification: $p = \text{Softmax}(\text{Matmul}(\hat{D}, \hat{\mathcal{C}}))$. The logits will then be used for the model optimization based on cross-entropy loss with the target of prediction task as:
\begin{equation}
\label{eq:pred}
    \mathcal{L}_{pred} = - \sum \text{log} P(\hat{c}_{l+1} | [BOS], I),
\end{equation}
where $\hat{c} = \text{argmax}(p)$ is the predicted code with highest probability.

\subsection{Feature Refinement} \label{subsec:rank}

\subsubsection{Enhanced Features
} In live-streaming recommendation system, the recommendation values between users and authors need to be predicted based on some probabilities (\textit{e.g.} click CTR, follow WTR, long view LVTR, gift GTR) given by a ranking model. The input of the ranking model is a series of features on both the user and author sides, 
and iterative refinement of these features enhances the model’s capability to deliver accurate recommendations. 

Considering the output results of the two key components of the prediction model: the output of encoder $\hat{I}$ is a dynamic fusion of the long-term segment semantics of the author, which also serves as the reflection of the author's content understanding; the output of decoder $\hat{D}$ is the prediction of the author's future content, which significantly influences the user's interaction behavior within the next segment interval. 
Therefore, we regard them as additional features passed into the ranking model:
\begin{equation}
\label{eq:rank}
    \hat{y}^{ctr}, \hat{y}^{wtr}, \dots = \text{MultiTask}(\text{Concat}[\text{uId}, \text{aId}, \text{sg}(\hat{I}), \text{sg}(\hat{D})]),
\end{equation}
\begin{equation}
    \mathcal{L}_{rank} = - \sum^{ctr, wtr, \dots}_{xtr}(y^{xtr}\text{log}\hat{y}^{xtr} + (1 - y^{xtr})\text{log}(1 - \hat{y}^{xtr})),
\end{equation}
where $\text{MultiTask}(\cdot)$ is a multi-gate mixture-of-experts model, and $\text{sg}(\cdot)$ means stop gradient, which aims at truncating the influence on the prediction model during the process of ranking. In this way, the ranking model is empowered to understand the recent content of the author, as well as to grasp user preferences based on foresight.

\subsubsection{Streaming Training}

The ranking stage in live-streaming recommendation system aims to score each author in the candidate sets for each user. Given that living segments are streaming data, we maintain a database to store the recent Sid sequence associated with each author. These Sids are derived from segments through the fine-tuned LLM and pre-trained K-Means to form discrete codes. 

During ranking, the historical Sid sequence is processed by the prediction model twice.
First, for model training, the last Sid is taken as target while the sequence of length $l$ before is the context, following the formulation in Eq.\ref{eq:pred}. 
Second, for next segment prediction, the same model is applied to the recent Sid sequence, and the results is then incorporated into the ranking model like Eq.\ref{eq:rank}. Figure \ref{fig:model} shows architecture of our model.

\section{Experiment}

\begin{table*}[ht!]
  \caption{A/B testing results in online live-streaming service platform.}
  \vspace{-0.3cm}
  \setlength{\tabcolsep}{2.5pt}
  \label{tab:ab}
  \begin{tabular}{ccccccccc}
    \toprule
    & \multicolumn{3}{c}{\textbf{Core Metrics}}  & \multicolumn{4}{c}{\textbf{Interaction Metrics}} 
    \\ \cmidrule(r){2-4} \cmidrule(r){5-8} & Exposure & Watch Count & Watch Time & Like & Follow & Gift Count & Gift Users \\
    \midrule
    \rowcolor{green!10}
    \multirow{2.5}{*}{\cellcolor{white}\textbf{Overall}} & \textbf{+0.457\%} & \textbf{+0.243\%} & \textbf{+0.370\%} & \textbf{+1.754\%} & \textbf{+0.922\%} & \textbf{+2.480\%} & \textbf{+1.011\%} \\ 
    & [-0.76\%, +1.71\%] & [+0.08\%, +0.40\%] & [-0.09\%, +0.82\%] & [+0.67\%, +2.80\%] & [-0.25\%, +2.15\%] & [+0.13\%, +4.91\%] & [+0.08\%, +1.93\%] \\ 
    \bottomrule
  \end{tabular}
  \vspace{-0.3cm}
\end{table*}

\subsection{Experimental Settings}

We evaluate our model at an industrial live-streaming recommendation scenario, which is one of the largest live-streaming platform serving hundreds of millions of users and millions of authors, with billions of interactions every day. 
We follow the widely-used evaluation metrics, \textit{i.e.} AUC and GAUC, to evaluate our model. 

\subsection{Overall Performance Experiments}

\subsubsection{Offline performance} We conduct an offline comparison experiment of gradually adding additional features produced from our model to the base ranking model. The results are shown in Table \ref{tab:offline}. We summarize the following observations: (i) Adding the encoded sequence information of the historical content $\hat{I}$ enhances the model's ability to understand the author. However, this information might have already been adequately modeled in the base ranking model, and thus only achieving limited results on some metrics. (ii) Further introduction of foresight prediction $\hat{D}$ achieves an incredible improvement in all metrics. This is because our method offers a new perspective, that is, to evaluate the possible interaction of users based on the future content of the authors, which is proved to be intuitive and effective. Besides, the more significant improvement of GAUC compared to AUC demonstrates the ability of our model to capture of the preference differences among various users.

\subsubsection{A/B testing performance} 

We deploy our model serving for ranking stage of live-streaming services platform to evaluate its real business contribution in an online A/B testing scenario. The experiment results shown in Table \ref{tab:ab} demonstrate the significant improvement in core metrics and interaction metrics, indicating the effective enhancement of user retention and consumption ability. 
These findings prove that modeling the dependency and changing trend of the author's content sequence can indeed bring real-time high-quality distribution to the live-streaming recommendation services and provide better user experience.

\begin{table}[t!]
  \caption{Prediction accuracy of different Strategies.}
  \vspace{-0.3cm}
  \label{tab:acc}
  \begin{tabular}{ccccccccc}
    \toprule
    \textbf{Strategy} & \textbf{Our Model} & Last\protect\footnotemark & Max Freq\protect\footnotemark & Max Weight\protect\footnotemark \\
    \midrule
    \textbf{Accuracy} & \cellcolor{green!10}\textbf{21.35\%} & 8.32\% & 8.94\% & 9.03\%  \\ 
    \bottomrule
  \end{tabular}
  \vspace{-0.3cm}
\end{table}

\footnotetext[1]{Choose last Sid $L_c[-1]$.} 
\footnotetext[2]{Choose max frequency in recent sequence $argmax(\sum_{i=0}^{l}\textbf{1}_{L_c[i] = x})$.} 
\footnotetext[3]{Choose max frequency in compressed sequence $argmax(\sum_{i=0}^{l}W_c[i]\times \textbf{1}_{L'_c[i] = x})$.}

\begin{figure}[t!]
\begin{center}
\includegraphics[width=7.5cm]{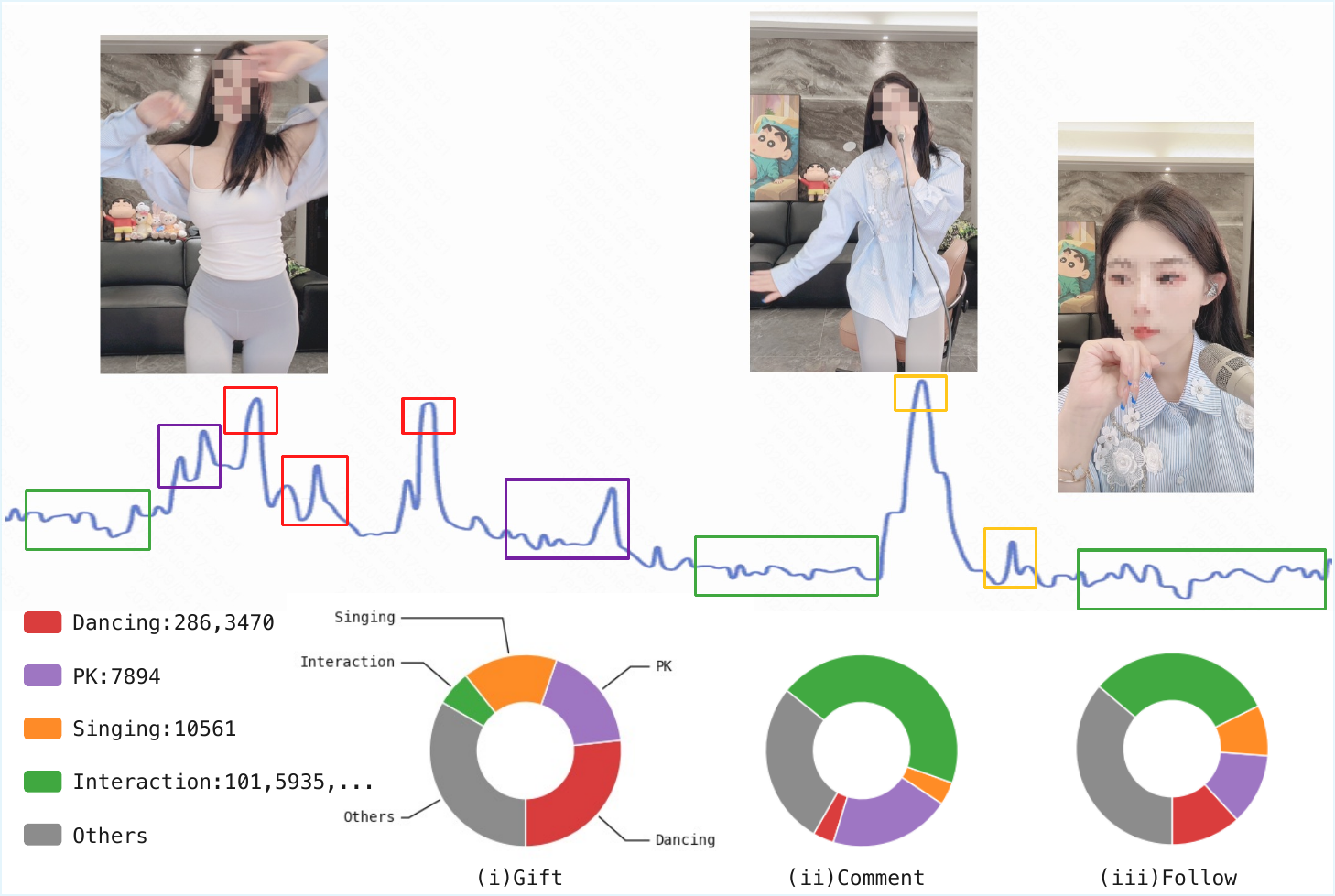}
\vspace{-0.2cm}
\caption{Analysis of Sid changes in a live-streaming case.}
\label{fig:cluster}
\end{center}
\vspace{-0.3cm}
\end{figure}

\subsection{Sid Analysis Experiments}

\subsubsection{Prediction accuracy}

To evaluate the accuracy of our model in predicting the subsequent content of the author, we compare our model-based prediction method with three rule-based methods, with the results shown in Table \ref{tab:acc}. 
It is clear that the model-based method provides a more accurate prediction of the next Sid, thereby improving the recommendation capability of ranking.

\subsubsection{Living Sid changes analysis}

We conduct a case study of Sid changes within a live-streaming to analyze its ability in precisely clustering of the author's content and correlation with user interaction, presented in Figure \ref{fig:cluster}. 
On the one hand, the segments like singing and dancing are all quantified to the same one or several Sids, demonstrating the excellent aggregation of content. 
On the other hand, through the calculation of the interaction proportion during the segment period, we find that some key Sids are strongly correlated with revenue indicators, \textit{e.g.}, user gifts mostly occur in the dancing and singing segments, while comments or follows take place within the interaction segments. 
These analyses can explain the performance gain brought about by content understanding.

\subsubsection{Case analysis} 
\begin{figure}[t!]
\begin{center}
\includegraphics[width=7.5cm]{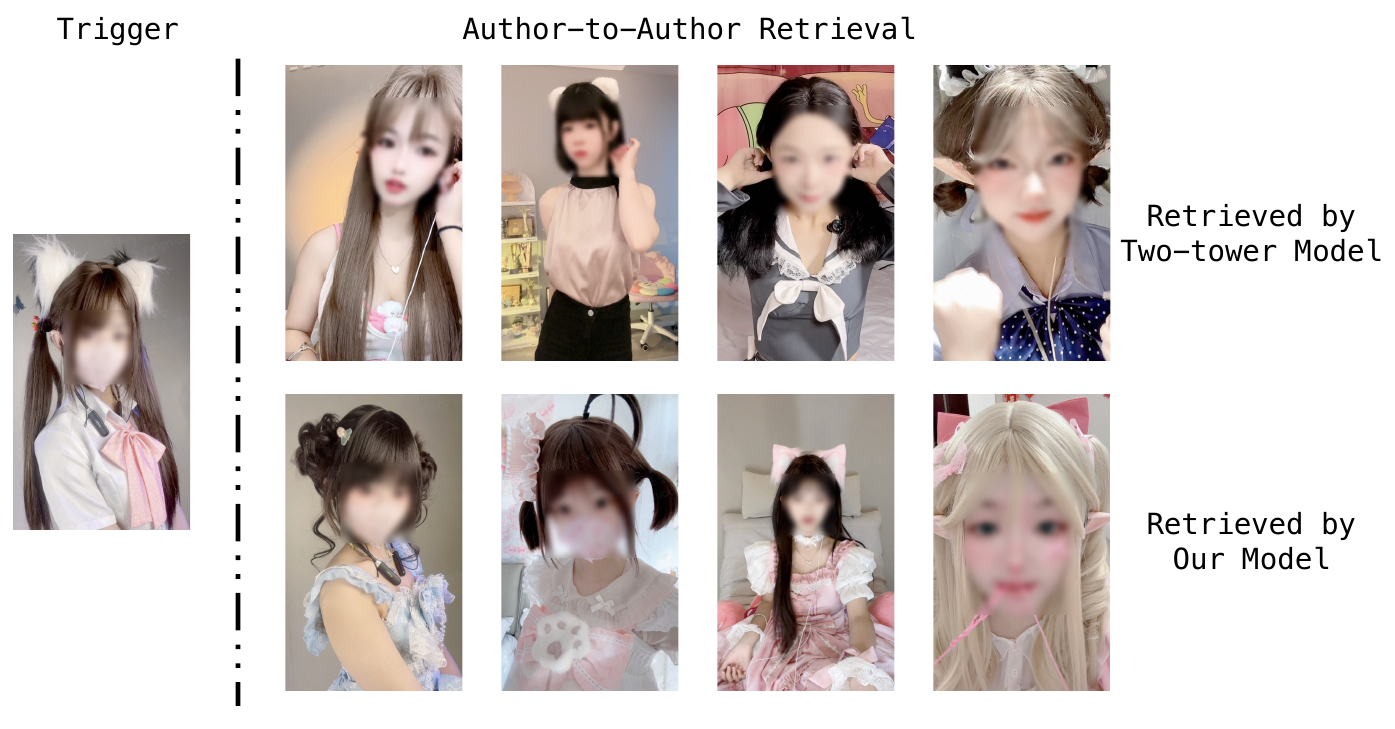}
\vspace{-0.2cm}
\caption{Case analysis of author-to-author retrieval. }
\label{fig:case}
\end{center}
\vspace{-0.5cm}
\end{figure}

To visualize the clustering effect of Sid, based on the data of users' interactions with the same behavior under the same Sid, we deploy an author-to-author retrieval service. Figure \ref{fig:case} shows one search result comparison case, where we can find that although the basic two-tower model can retrieve visual-appeal  authors, our model can retrieve more fine-grained related Lolita-fashion authors, which can be attributed to the multi-modal perception introduced by Sid. Moreover, considering the data flow for service construction, users' interaction preferences show a certain degree of consistency and continuity. Therefore, it is feasible to evaluate users' retention by predicting authors' future content.

\section{Conclusion}

In this paper, we propose an effective method of integrating foresight prediction into live-streaming recommendation system to enhance the model's perception of content changes. We discretize live-streaming content into semantic ids through quantization, and subsequently encode historical segment sequence for overall understanding and decode to predict future content. Extensive offline and online experiments demonstrate the effectiveness of our method.

\balance
\bibliographystyle{ACM-Reference-Format}
\bibliography{sample-base-extend.bib}

@article{moment,
  title={Moment\&Cross: Next-Generation Real-Time Cross-Domain CTR Prediction for Live-Streaming Recommendation at Kuaishou},
  author={Cao, Jiangxia and Wang, Shen and Li, Yue and Wang, Shenghui and Tang, Jian and Wang, Shiyao and Yang, Shuang and Liu, Zhaojie and Zhou, Guorui},
  journal={arXiv preprint arXiv:2408.05709},
  year={2024}
}

@inproceedings{mmbee,
  title={MMBee: Live Streaming Gift-Sending Recommendations via Multi-Modal Fusion and Behaviour Expansion},
  author={Deng, Jiaxin and Wang, Shiyao and Wang, Yuchen and Qi, Jiansong and Zhao, Liqin and Zhou, Guorui and Meng, Gaofeng},
  booktitle={Proceedings of the 30th ACM SIGKDD Conference on Knowledge Discovery and Data Mining},
  pages={4896--4905},
  year={2024}
}

@article{farm,
  title={FARM: Frequency-Aware Model for Cross-Domain Live-Streaming Recommendation},
  author={Li, Xiaodong and Yang, Ruochen and Wen, Shuang and Wang, Shen and Liu, Yueyang and Wang, Guoquan and Hu, Weisong and Luo, Qiang and Sheng, Jiawei and Liu, Tingwen and others},
  journal={arXiv preprint arXiv:2502.09375},
  year={2025}
}

@article{contentctr,
  title={ContentCTR: Frame-level live streaming click-through rate prediction with multimodal transformer},
  author={Deng, Jiaxin and Shen, Dong and Wang, Shiyao and Wu, Xiangyu and Yang, Fan and Zhou, Guorui and Meng, Gaofeng},
  journal={arXiv preprint arXiv:2306.14392},
  year={2023}
}

@inproceedings{tsstfn,
  title={A Bilateral Perspective for Modeling Real-Time Traffic Trends in Live-Streaming Recommendation},
  author={Li, Rui and Gao, Pengyuan and Li, Haihan and Chai, Ling and Huang, Shaohao and Xie, Ting},
  booktitle={2025 IEEE 41st International Conference on Data Engineering (ICDE)},
  pages={4142--4155},
  year={2025},
  organization={IEEE}
}

@article{larm,
  title={LLM-Alignment Live-Streaming Recommendation},
  author={Liu, Yueyang and Cao, Jiangxia and Wang, Shen and Wen, Shuang and Chen, Xiang and Wu, Xiangyu and Yang, Shuang and Liu, Zhaojie and Gai, Kun and Zhou, Guorui},
  journal={arXiv preprint arXiv:2504.05217},
  year={2025}
}

@article{liveforesighter,
  title={LiveForesighter: Generating Future Information for Live-Streaming Recommendations at Kuaishou},
  author={Lu, Yucheng and Cao, Jiangxia and Kuan, Xu and Cheng, Wei and Jiang, Wei and Zhang, Jiaming and Shuang, Yang and Zhaojie, Liu and Hong, Liyin},
  journal={arXiv preprint arXiv:2502.06557},
  year={2025}
}

@article{vqvae,
  title={Neural discrete representation learning},
  author={Van Den Oord, Aaron and Vinyals, Oriol and others},
  journal={Advances in neural information processing systems},
  volume={30},
  year={2017}
}

@article{tiger,
  title={Recommender systems with generative retrieval},
  author={Rajput, Shashank and Mehta, Nikhil and Singh, Anima and Hulikal Keshavan, Raghunandan and Vu, Trung and Heldt, Lukasz and Hong, Lichan and Tay, Yi and Tran, Vinh and Samost, Jonah and others},
  journal={Advances in Neural Information Processing Systems},
  volume={36},
  pages={10299--10315},
  year={2023}
}

@article{onerec,
  title={Onerec: Unifying retrieve and rank with generative recommender and iterative preference alignment},
  author={Deng, Jiaxin and Wang, Shiyao and Cai, Kuo and Ren, Lejian and Hu, Qigen and Ding, Weifeng and Luo, Qiang and Zhou, Guorui},
  journal={arXiv preprint arXiv:2502.18965},
  year={2025}
}

@inproceedings{fm,
  title={Factorization machines},
  author={Rendle, Steffen},
  booktitle={2010 IEEE International conference on data mining},
  pages={995--1000},
  year={2010},
  organization={IEEE}
}

@inproceedings{din,
  title={Deep interest network for click-through rate prediction},
  author={Zhou, Guorui and Zhu, Xiaoqiang and Song, Chenru and Fan, Ying and Zhu, Han and Ma, Xiao and Yan, Yanghui and Jin, Junqi and Li, Han and Gai, Kun},
  booktitle={Proceedings of the 24th ACM SIGKDD international conference on knowledge discovery \& data mining},
  pages={1059--1068},
  year={2018}
}

@article{dmcdr,
  title={Exploring Preference-Guided Diffusion Model for Cross-Domain Recommendation},
  author={Li, Xiaodong and Tang, Hengzhu and Sheng, Jiawei and Zhang, Xinghua and Gao, Li and Cheng, Suqi and Yin, Dawei and Liu, Tingwen},
  journal={arXiv preprint arXiv:2501.11671},
  year={2025}
}
\end{document}